\documentclass[a4paper,11pt]{article}
\usepackage{jheppub}
\usepackage[T1]{fontenc}
\usepackage{amsfonts}
\usepackage{amsmath}
\usepackage{amssymb,epsf}
\usepackage{latexsym}
\usepackage{graphicx,epsfig}
\usepackage{amssymb}
\usepackage{subfigure}

\title{Holographic Superconductors with Logarithmic Nonlinear Electrodynamics in an External Magnetic Field}

\author[a,b]{A. Sheykhi}
\author[a]{F. Shamsi}
\affiliation[a]{ Physics Department and Biruni Observatory,
College of Sciences, Shiraz University, Shiraz 71454, Iran}
\affiliation[b]{Research Institute for Astronomy and Astrophysics
of Maragha (RIAAM), P.O. Box 55134-441, Maragha, Iran}
\emailAdd{asheykhi@shirazu.ac.ir}

\abstract{ Based on the matching method, we explore the effects of
adding an external magnetic field on the $s$-wave holographic
superconductor when the gauge field is in the form of the
logarithmic nonlinear source. First, we obtain the critical
temperature as well as the condensation operator in the presence
of logarithmic nonlinear electrodynamics and understand that they
depend on the nonlinear parameter $b$. We show that the critical
temperature decreases with increasing $b$, which implies that the
nonlinear gauge field makes the condensation harder. Then, we turn
on the magnetic field in the bulk and find the critical magnetic
field, $B_c$, in terms of the temperature, which also depends on
the nonlinear parameter $b$. We observe that for temperature
smaller than the critical temperature, $T<T_c$, the critical
magnetic field increases with increasing $b$ and goes to zero as
$T\rightarrow T_c$, independent of the nonlinear parameter $b$. In
the limiting case where $ b\rightarrow0 $, all results restore
those of the holographic superconductor with magnetic field in
Maxwell theory.}
\begin{document}

\maketitle
\section{Introduction}
The most successful microscopic theory of superconductivity was
proposed by Bardeen, Cooper and Schrieffer (BCS) \cite{BCS}. The
BCS theory describes various properties of usual (low temperature)
superconducting materials with great accuracy. The gauge/gravity
duality \cite{Mal,Wit} states that string theory in asymptotic
anti-de Sitter (AdS) spacetime can be dual to a conformal field
theory (CFT) living on its boundary. This duality provides a
well-established method for calculating the properties of the
superconductors using a dual classical gravity description. The
mechanism of the high temperature superconductors has long been an
unsolved mysteries in modern condensed matter physics. Much
research is now directed towards solving such worried by helping
of different gravitational systems.

A great step toward understanding the strong coupling
superconductivity put forwarded by Hartnoll, et. al., in $2008$
\cite{Har1,Har2} who disclosed that some properties of strongly
coupled superconductors can be potentially described by classical
general relativity living in one higher dimension. Such strongly
coupled superconducting phases of the boundary field theory are
termed \textit{holographic superconductors} in the literatures.
The construction described above relates to a $s$-wave holographic
superconductor and typically involves an Abelian Higgs model with
a bulk complex scalar field that is charged under the Maxwell
field \cite{Hor}. According to the ADS/CFT correspondence, in the
gravity side, a Maxwell field and a charged scalar field are
introduced to describe the $U(1)$ symmetry and the scalar operator
in the dual field theory, respectively. This holographic model
undergoes a phase transition from black hole with no hair (normal
phase/conductor phase) to the case with scalar hair at low
temperatures (superconducting phase). Various aspects of the
holographic superconductors have been explored from different
perspective \cite{Mus,RGC1,P.BGRL, P.MRM, P.CW,
P.ZGJZ,RGC2,Wang2,Wang3,BI-GB-BR,Shey}.

It is also interesting to investigate the effects of an external
magnetic field on the holographic superconductors. One of the
major properties of ordinary superconductors is that they exhibit
perfect diamagnetism as the temperature is lowered below $T_c$ in
the presence of an external magnetic field. In other words, at low
temperature, superconductors expel magnetic field line and the
phenomenon is known as Meissner effect. It is worthy to explore
whether or not such an effect can be seen in the holographic
superconductors when the magnetic field is turned on. Some efforts
have been done to disclose the properties of the holographic
superconductors in the presence of an external magnetic field
using both numerical approaches as well as analytical analysis
\cite{Mon, Mag, Al,Ma}. As an analytical approach for deriving the
upper critical magnetic field, an expression was found in the
probe limit by extending the matching method first proposed in
\cite{Gr} to the magnetic case \cite{XH}, which is shown to be
consistent with the Ginzburg-Landau theory.

Most of these analysis are carried out in the framework of Maxwell
electrodynamics. Along with the conventional Maxwell
electrodynamic theory, nonlinear electrodynamic theories, which
correspond to the higher derivative corrections to the Abelian
gauge fields, have also become interesting topics of research in
the past several decades. The primary motivation for introducing
nonlinear electrodynamics was to remove the divergences in the
self-energy of point-like charged particles \cite{BI}. However,
they have earned renewed attention over past several years since
these theories naturally arise in the low-energy limit of the
heterotic string theory \cite{Ka,RGC}. Recently, the response of a
holographic superconductor to an external magnetic field in the
presence of Born-Infeld nonlinear electrodynamics was investigated
by applying a simple analytical approach based on the matching of
the solutions to the field equations near the horizon and near the
asymptotic AdS region \cite{Gan}. Unfortunately, the final result
obtained for the critical magnetic field in Eq. (48) of Ref.
\cite{Gan}  is not correct, unless for the Maxwell case where
$b=0$. Indeed, one can easily check that $B_c$ derived in
\cite{Gan} goes to zero as temperature tends to zero, while we
expect $B_c$ approaches a finite value as $T$ goes to zero.
Besides, $B_c$ of Ref. \cite{Gan} does not goes to zero as
$T\rightarrow T_c$, while according to the definition, we should
have $B_c\rightarrow0$  in the limiting case where $T\rightarrow
T_c$.  Using the matching method in the probe limit, some
properties of holographic superconductor in Gauss-Bonnet gravity
with Born-Infeld electrodynamics and in the presence of a critical
magnetic field were investigated in \cite{Xue}. Performing
explicit analytic computations, with and without magnetic field,
the effects of higher order corrections to gravity as well as
gauge field on the holographic s-wave condensation and
Meissner-like effect have been explored in \cite{Lala, Roy}. In
recent years, other types of nonlinear electrodynamics in the
context of gravitational field have also been introduced, which
can also remove the divergence of the electric field at the
origin, similar to Born-Infeld nonlinear electrodynamics. Two
well-known nonlinear Lagrangian for electrodynamics are
logarithmic \cite{Soleng} and exponential \cite{HendiJHEP}
Lagrangian. In this paper, we shall provide the matching method to
investigate the effects of the logarithmic nonlinear
electrodynamics (LNE) on the holographic superconductors with an
external magnetic field. The case with exponential nonlinear
electrodynamics will be addressed in our future investigation.

This paper is organized as follows. In the next section, we
introduce the basic field equations of holographic superconductors
with LNE and obtain the critical temperature as well as the
critical exponent. In section III, we turn on an external magnetic
field and investigate the effects of it as well as nonlinear
corrections to the gauge field on the properties of the
holographic superconductors. In particular, we obtain the critical
magnetic field in terms of the temperature which is affected by
the nonlinear gauge field. We finish our paper with conclusions
and discussions in section IV.

\section{Basic Equations of Holographic Superconductors with LNE\label{Basic}}
Our starting point is a planar Schwarzschild AdS black holes,
which is described by the line element
\begin{eqnarray}\label{metric}
ds^{2}&=- f(r) dt^{2}+\frac{dr^2}{f(r)}+r^{2}(dx^2 + dy^2),
\end{eqnarray}
where
\begin{equation}
f(r) = r^2- \frac{r_{+}^3}{r},
\end{equation}
in units in which the AdS radius is unity, i.e. $ l=1$. In the
above relation $ r_{+} $ is  the horizon radius. The Hawking
temperature associated with the horizon is
\begin{equation}\label{T}
T =\frac{f^{\prime}(r_{+})}{4\pi}=\frac{3 r_{+}}{4 \pi}.
\end{equation}
Let us now consider an electric field and a charged complex scalar
field in this fixed background. The corresponding Lagrangian
density can be expressed as
\begin{eqnarray}\label{lagrangian}
\mathcal{L}=\mathcal{L}_{LN}- | \nabla_{\mu}\psi - i q A_{\mu}
\psi|^{2} - m^{2}|\psi|^{2},
\end{eqnarray}
where $ A_{\mu} $ and $ \psi $ are the gauge and scalar field,
respectively. The term $ \mathcal{L}_{LN} $ in (\ref{lagrangian})
corresponds to the Lagrangian of the logarithmic nonlinear
electrodynamics, which we take it in the following form
\cite{Soleng}
\begin{eqnarray}\label{La}
\mathcal{L}_{LN}=-\frac{1}{b} \ln\bigg(1+\frac{b F}{4}\Bigg),
\end{eqnarray}
where $F \equiv F_{\mu\nu}F^{\mu\nu}$ and $F^{\mu\nu}$ is the
electromagnetic field tensor. The constant $b$ is the nonlinear
parameter which indicates the strength of the nonlinearity. In the
limiting case where $b\rightarrow0$, the Lagrangian $
\mathcal{L}_{LN} $ reduces to the standard Maxwell Lagrangian,
$\mathcal{L}_{M}=-\frac{1}{4}F_{\mu\nu}F^{\mu\nu}$. We consider
the following ansatz for the  gauge and scalar fields \cite{Har1}
\begin{eqnarray}
A_{\mu}=\left(\phi(r),0,0,0\right),  \quad  \quad \psi=\psi(r),
\end{eqnarray}
Varying Lagrangian (\ref{lagrangian}) with respect to $\phi(r)$
and $\psi(r)$, the scalar and gauge field equations become
\begin{eqnarray}\label{phir}
\bigg(2+ b\phi'^{2}(r)\bigg)\phi^{\prime\prime}(r) +\frac{2}{r}
\bigg(2- b \phi'^{2}(r)\bigg)\phi'(r) -\frac{\phi(r)}{f(r)}
\psi^{2} (r) \bigg(2- b \phi'^{2}(r)\bigg)^{2}=0,\label{Eqphi}
\end{eqnarray}
\begin{eqnarray}\label{psir}
\psi^{\prime\prime}(r) +\left(\frac{f'}{f}+\frac{2}{r}\right)
\psi'(r) +\bigg(\frac
{\phi^2(r)}{f^2}-\frac{m^2}{f}\bigg)\psi(r)=0,
\end{eqnarray}
where prime denotes derivative with respect to $r$. One may note
that for $ b\rightarrow 0 $, the field equation (\ref{phir}) for
gauge field, restore the field equation of holographic
superconductor in Maxwell theory \cite{Har2}. In order to solve
the non-linear equations (\ref{phir}) and (\ref{psir}), we should
find the boundary condition for $ \phi $ and $ \psi $ near the
black hole horizon $ r \sim r_{+} $ as well as at the spatial
infinite $ r \rightarrow\infty$. Using Eq. (\ref{psir}) and the
fact $ f(r_{+})=0 $ and $f'(r_{+})=3 r_{+} $, one can show that
the regularity condition at the horizon leads to the boundary
conditions $ \phi(r_{+})=0$ and $ \psi'(r_{+})=-\frac{2}{3 r_{+}}
\psi(r_{+}) $.

Next, we transform the coordinate as $r\rightarrow
z=\frac{r_{+}}{r} $ and also set $ m^2=-2 $. In the new
coordinate,  the horizon ($ r=r_{+} $ ) and the boundary ($
r\rightarrow \infty $), are correspond to $ z = 1 $ and $ z = 0 $,
respectively. The equations of motion for the scalar field $
\psi(z) $ and the gauge field $ A_{\mu}$ translate to
\begin{eqnarray}\label{phiz}
\bigg(2+ b \frac{z^4}{r_{+}^2}\phi'^2(z)
\bigg)\phi^{\prime\prime}(z) + \frac{4b z^{3} }{r_{+} ^{2}}
{{\phi^{\prime}}^{3}(z)}
-\frac{r_{+}^2}{z^4}\frac{\phi(z)}{f(z)}\psi^2(z)\bigg(2- b
\frac{z^4}{r_{+}^2}\phi'^2(z)\bigg)^2=0,
\end{eqnarray}
\begin{eqnarray}
\psi^{\prime\prime}(z)+\frac{f^{\prime}(z)}{f(z)}
\psi^{\prime}(z)+\frac{2 r_{+}^{2}}{z^4 f(z)}
\psi(z)+\frac{r_{+}^2\phi^2(z)}{z^4 f^2(z)} \psi(z)=0,
\end{eqnarray}
where prime now denotes derivative with respect to $ z $ and $
f'(1)=-3 r_{+}^2 $. Let us now take a look at the boundary
conditions in the new coordinate $z$. It is easy to check that the
regularity at the horizon $z=1$ implies,
\begin{equation}
\phi(1)=0,\quad \psi'(1) =\frac{2}{3}\psi(1).
\end{equation}
While, in the asymptotic region ($ z\rightarrow 0 $) the solutions
may be written as,
\begin{equation}\label{limphi}
\phi(z)\thickapprox\mu-\frac{\rho}{r_{+}} z,\quad
\psi(z)\thickapprox J_{-} z+J_{+} z^2,
\end{equation}
where $ \mu $ and $ \rho $ are interpreted as a chemical potential
and charge density, respectively. In what follows we set $ J_{+} =
0 $. The condensation operator $ <\mathcal{O}_{-}> $ is related to
$ J_{-} $ as $ <\mathcal{O}_{-}> = \sqrt{2} r_{+}J_{-} $
\cite{Gan}.

Now we are in a position to obtain expressions for the critical
temperature and values of the condensation, analytically. Let us
first consider the solutions of the gauge field, $ \phi(r) $, and
the scalar field, $ \psi(z) $, near the horizon ($ z = 1 $). Using
the Taylor expansion for $ \phi(z) $ and  $ \psi(z) $ near the
horizon, we can write
\begin{eqnarray}\label{taylorphi}
\phi(z) &=& \phi(1)-\phi'(1) (1-z)+\frac{1}{2} \phi''(1) (1-z)^2+
... \nonumber \\ &\thickapprox& -\phi'(1) (1-z)+\frac{1}{2}
\phi''(1) (1-z)^2,
\end{eqnarray}
\begin{eqnarray}\label{taylorpsi}
\psi(z)=\psi(1)-\psi'(1)(1-z)+\frac{1}{2}\psi''(1)(1-z)^2+...,
\end{eqnarray}
where we have used the boundary condition $ \phi(1) = 0 $ in the
second line of Eq. (\ref{taylorphi}). We further assume $ \phi'(1)
< 0 $ and $ \psi(1) > 0 $ in order to make $ \phi(z) $ and $
\psi(z) $ positive. This can be done without any loss of
generality. Near the horizon, $z=1 $, and using the relations $
f'(1)=-3 r_{+}^2 $ and $ f''(1) =6r_{+}^2$, one can show that
\begin{eqnarray}\label{phi1}
\phi''(1)=-\frac{2b \phi'^3(1)}{r_{+}^2} -\frac{2}{3} \phi'(1)
\psi^2(1)
\bigg(1-\frac{3b}{2r_{+}^2}\phi'^2(1)\bigg)+\mathcal{O}(b^2),
\end{eqnarray}
\begin{eqnarray}\label{psi1}
\psi''(1)=-\frac{4}{9}\psi(1)-\frac{\phi'^2(1)\psi(1)}{18
r_{+}^2}.
\end{eqnarray}
where we have only kept the expression up to the linear term of
the nonlinear parameter $b$, by assuming that $b$ is small.
Finally, substituting Eqs. (\ref{phi1}) and (\ref{psi1}) in the
Taylor expansion (\ref{taylorphi}), (\ref{taylorpsi}), one gets
\begin{eqnarray}\label{m.phi}
\phi(z)\thickapprox-\phi'(1)(1-z)-\bigg[\frac{b}{r_{+}^2}\phi'^2(1)+\frac{1}{3}\psi^2(1)\bigg(1-\frac{3b}{2r_{+}^2}\phi'^2(1)\bigg)\bigg]\phi'(1)(1-z)^2+\mathcal{O}(b^2),
\end{eqnarray}
\begin{eqnarray}\label{m.psi}
\psi(z)&\thickapprox&\psi(1)-\psi'(1)(1-z)-\bigg[\frac{2}{9}\psi(1)+\frac{\phi'^2(1)\psi(1)}{36
r_{+}^2}\bigg](1-z)^2\nonumber\\ &=&
 \frac{1}{3}\psi(1)+\frac{2}{3}\psi(1)z - \bigg[\frac{2}{9}\psi(1)+\frac{\phi'^2(1)\psi(1)}{36
 r_{+}^2}\bigg](1-z)^2,
\end{eqnarray}
where in the last step of (\ref{m.psi}), we have also used the
regularity conditions for $\psi$ at the horizon $z=1$. Now, using
the method prescribed by the matching technique \cite{Gr}, we
match the solutions (\ref{limphi}) with solutions (\ref{m.phi})
and (\ref{m.psi}) at some intermediate point $ z = z_{m} $. It is
easy to check that the matching of the two asymptotic solutions
smoothly at $ z = z_{m} $ lead to the following four conditions
\begin{eqnarray}
\mu-\frac{\rho}{r_{+}}z_{m}=\beta(1-z_{m})+\bigg[\frac{b}{r_{+}^2}\beta^2+\frac{1}{3}\alpha^2\bigg(1-\frac{3b}{2r_{+}^2}\beta^2\bigg)\bigg]\beta(1-z_{m})^2,
\end{eqnarray}
\begin{eqnarray}\label{eq20}
\frac{\rho}{r_{+}}=\beta+2\beta\bigg[\frac{b}{r_{+}^2}\beta^2+\frac{1}{3}\alpha^2\bigg(1-\frac{3b}{2r_{+}^2}\beta^2\bigg)\bigg](1-z_{m}),
\end{eqnarray}
\begin{eqnarray}\label{eq21}
J_{-}z_{m}=\frac{\alpha}{3}+\frac{2\alpha}{3}z_{m}-\frac{\alpha}{9}\bigg[2+\frac{\beta^2}{4r_{+}^2}\bigg](1-z_{m})^2,
\end{eqnarray}
\begin{eqnarray}\label{eq22}
J_{-}=\frac{2\alpha}{3}+\frac{\alpha}{9}\bigg[4+\frac{\beta^2}{2r_{+}^2}\bigg](1-z_{m}),
\end{eqnarray}
where we have defined $\alpha\equiv\psi(1)$, $
\beta\equiv-\phi'(1) $ ($ \alpha, \beta > 0 $). From Eq.
(\ref{eq20}), we find
\begin{eqnarray}
\alpha^2=\frac{3}{2(1-z_{m})}\bigg[\frac{\rho}{\tilde{\beta}r_{+}^2}+\frac{3\rho
b \tilde{\beta}}{2r_{+}^2}-\frac{7}{2}b\tilde{\beta}^2+2b
\tilde{\beta}^2 z_{m}-1 \bigg]+\mathcal{O}(b^2),
\end{eqnarray}
which by using (\ref{T}), can be rewritten
\begin{eqnarray}\label{alpha2}
\alpha^2=\frac{3}{2(1-z_{m})}\left[1+\frac{b\tilde{\beta}^2}{2}(7-4z_{m})\right]\frac{T_{c}^2}{T^2}\left(1-\frac{T^2}{T_{c}^{2}}\right)+\mathcal{O}(b^2),
\end{eqnarray}
where $ \tilde{\beta}={\beta}/{r_{+}}$ and the quantity $ T_{c} $
may be identified as the critical temperature for condensation and
is given by,
\begin{equation}\label{Tc}
T_{c}=\kappa \sqrt{\rho},
\end{equation}
where
\begin{equation}\label{kappa}
\kappa=\frac{3}{4
\pi\sqrt{\tilde{\beta}}}\sqrt{1-2b\tilde{\beta}^2(1-z_{m})}.
\end{equation}
When the temperature is very close to the critical temperature ($
T \sim T_{c} $), from Eq. (\ref{alpha2}) we obtain,
\begin{eqnarray}\label{eq26}
\alpha=\sqrt{\frac{3}{(1-z_{m})}}\left(1+\frac{b\tilde{\beta}^2}{4}(7-4z_{m})
\right)\sqrt{1-\frac{T}{T_{c}}}+\mathcal{O}(b^2).
\end{eqnarray}
It is worth noting that in the expressions for $\alpha$ and
$T_{c}$  we again kept up to the linear terms in $b$. From Eqs.
(\ref{eq21}), (\ref{eq22}), we obtain
\begin{eqnarray}\label{eq27}
\tilde{\beta}=2 \sqrt{\frac{1+2z_{m}^2}{1-z_{m}^2}},\quad
J_{-}=\frac{2 \alpha (2+z_{m})}{3(1+z_{m})}.
\end{eqnarray}
Note that one may arrive at the above value for $ \tilde{\beta} $
by using the fact that $ J_{+} =0$. Indeed, from
Eq.(\ref{limphi}), we have
\begin{equation}
\psi(z\rightarrow 0)\thickapprox\psi_{0} = J_{-} z+J_{+} z^2,
\end{equation}
which leads to
\begin{equation}
J_+=-\frac{\psi_0}{z^2}+\frac{\psi_0'}{z}=0.
\end{equation}
Combining the above condition with
\begin{eqnarray}
\psi_0=\frac{\alpha}{3}+\frac{2\alpha}{3}z_{m}-\frac{\alpha}{9}\bigg[2+\frac{\beta^2}{4r_{+}^2}\bigg](1-z_{m})^2,
\end{eqnarray}
we arrive at value (\ref{eq27})  for $ \tilde{\beta} $. Now, we
compute the condensation operator from Eqs. (\ref{eq26}),
(\ref{eq27}),
\begin{eqnarray}
<\mathcal{O}_{-}>=\sqrt{2} r_{+}J_{-}=\gamma T_{c}
\sqrt{1-\frac{T}{T_{c}}},
\end{eqnarray}
where we have defined
\begin{eqnarray}
\gamma=\frac{8
\sqrt{2}\pi(2+z_{m})}{9(1+z_{m})}\sqrt{\frac{3}{(1-z_{m})}\bigg(\frac{7}{2}b\tilde{\beta}^2-2b
\tilde{\beta}^2 z_{m}+1\bigg)}.
\end{eqnarray}
In is notable to mention that there is an upper bound on the
nonlinear parameter $b$. Indeed, from Eq. (\ref{kappa}) we see
that the nonlinear parameter should be satisfied in the following
upper bound
\begin{equation}\label{bbound}
b<\frac{1}{2
\tilde{\beta}^2(1-z_{m})}=\frac{1+z_{m}}{8(1+2z_{m}^2)}.
\end{equation}
Since  $0\leq z_m\leq1$, therefore for $z_{m}=0$, we have for the
upper bound $b<1/8$, and for $z_{m}=1$ we have $b<1/12$. In the
limiting case where $b\rightarrow0$, all the above relations
reduce to the results of holographic superconductor in external
magnetic field in Maxwell theory \cite{XH}. Let us summarize the
result for critical temperature in table $1$. We see that the
critical temperature decrease with increasing $ b $, which implies
that the nonlinear electrodynamics makes the condensation harder.
This is consistent with the results of \cite{Gan,GanGB}. Comparing
the results obtained in this table from matching method, with
those obtained from the numerical method \cite{Zh} and those
obtained from Sturm-Liouville variational method for Born-Infeld
nonlinear electrodynamics \cite{GanBI} shows that the
Sturm-Liouville method yields better results than the matching
method.
\begin{center}
\begin{tabular}{|c|c|c|c|c|c|c|}
\hline
$ b $\quad &   $0$\quad &   0.01\quad  &   0.02\quad  &   0.03\quad  &   0.04\quad  &   0.05\quad \\
\hline
$T_c/\sqrt{\rho}$ \quad &   0.168 \quad &   0.161\quad  &   0.155\quad  &   0.148\quad &   0.141 \quad &   0.133 \quad  \\
\hline
\end{tabular}
\\[0pt]
Table $1$: Critical temperature for the LN holographic
superconductors by using the matching method for different values
of nonlinear parameter $b$ at $ z_{m}=0.1 $. \label{tab1}
\end{center}
\section{Effects of an External Magnetic Field with LNE}
In this section we would like to investigate the effects of an
external magnetic field on the holographic superconductors in the
presence of LNE. For this purpose, we consider a magnetic field in
the bulk. The asymptotic value of this magnetic field corresponds
to a magnetic field added to the boundary field theory by the
gauge gravity correspondence. This allows us to adopt the
following ansatz for the gauge field and the scalar field
\cite{Al}
\begin{eqnarray}
A_{t}=\phi(z), \quad A_{y}=B x, \quad A_{x}=A_{z}=0, \quad
\psi=\psi(x,z).
\end{eqnarray}
Therefore, the equation for the scalar field $ \psi $ becomes
\begin{eqnarray}\label{psixz}
\psi''(x,z)+\frac{f'(z)}{f(z)} \psi'(x,z) + \frac{2 r_{+}^2
\psi(x,z)}{z^4 f(z)} + \frac{r_{+}^2 \phi^2(z)\psi(x,z)}{z^4
f^2(z)}+\frac{1}{z^2 f(z)}(\partial_{x}^2 {\psi} - B^2 x^2
\psi)=0.
\nonumber \\
\end{eqnarray}
In order to solve Eq. (\ref{psixz}), we shall use the method of
separation of variables. Let us consider the solution of the
following form
\begin{equation}\label{eq32}
\psi(x,z)=X(x) R(z).
\end{equation}
Substituting (\ref{eq32}) into Eq. (\ref{psixz}), we obtain the
following equation
\begin{eqnarray}\label{eq33}
z^2 f(z)\bigg[\frac{R''}{R}+\frac{f'}{f}
\frac{R'}{R}+\frac{r_{+}^2\phi^2}{z^4f^2}+\frac{2 r_{+}^2}{z^4
f}\bigg]-\bigg[-\frac{X''}{X}+B^2x^2\bigg]=0.
\end{eqnarray}
Clearly, the equations for two variables, $ x $ and $ z $ are
separable. The equation for $ X (x) $ is readily identified as the
Schrodinger equation in one dimension with a $ B $ dependent
frequency \cite{Al}
\begin{eqnarray}
-X''(x)+B^2x^2X(x)=\lambda_{n} B X(x),
\end{eqnarray}
where $ \lambda_{n}=2n+1 $ ($ n=\rm integer $). Since the most
stable solution corresponds to $ n = 0 $, the $ z $ dependent part
of (\ref{eq33}) may be expressed as \cite{Al}
\begin{eqnarray}\label{eq35}
R''(z)+\frac{f'(z)}{f(z)} R'(z)+\frac{2 r_{+}^2 R(z)}{z^4
f(z)}+\frac{r_{+}^2\phi^2(z)R(z)}{z^4f^2(z)}=\frac{BR(z)}{z^2
f(z)}.
\end{eqnarray}
Now at the horizon, $ z = 1 $, using (\ref{eq35}) and $ f'(1) =3
r_{+}^2$, we can write the following equation
\begin{equation}\label{eq36}
R'(1)=\bigg(\frac{2}{3}-\frac{B}{3 r_{+}^2}\bigg)R(1).
\end{equation}
On the other hand, at the asymptotic infinity, $ z\rightarrow0 $,
the solution of Eq. (\ref{eq35}) can be written as
\begin{equation}\label{limr}
R(z)=J_{-}z+J_{+}z^2.
\end{equation}
Note that in our analysis we shall choose $ J_{+} = 0 $. Our aim
here is to find the value of the (critical) magnetic field for
which the condensation vanishes. Again, we employ the matching
method \cite{Gr}. The Taylor expansion of $ R(z) $ around $z=1$
leads to
\begin{eqnarray}\label{eq38}
R(z)=R(1)-R'(1)(1-z)+\frac{1}{2}R''(1)(1-z)^2+...
\end{eqnarray}
Considering Eq. (\ref{eq35}) near the horizon ($z=1$) and using
the fact that $ f'(1)=-3 r_{+}^2 $, $ f''(1)=6r_{+}^2 $ as well as
the regularity condition for $ \phi $ and Eq. (\ref{eq36}), we
arrive at
\begin{eqnarray}\label{eq39}
R''(1)=\bigg[-\frac{4}{9}-\frac{2B}{9 r_{+}^2}+\frac{B^2}{18
r_{+}^4}-\frac{\phi'^2(1)}{18 r_{+}^2}\bigg]R(1),
\end{eqnarray}
Combining Eq. (\ref{eq39}) with Eqs. (\ref{eq36}) and
(\ref{eq38}), we find the following expression
\begin{eqnarray} \label{Rz}
R(z)\thickapprox\frac{1}{3}R(1)+\frac{2}{3}R(1)z+\frac{BR(1)}{3
r_{+}^2}(1-z)+\frac{1}{2}\bigg[-\frac{4}{9}-\frac{2B}{9
r_{+}^2}+\frac{B^2}{18 r_{+}^4}-\frac{\phi'^2(1)}{18
r_{+}^2}\bigg]R(1)(1-z)^2.
\end{eqnarray}
Following the arguments of matching technique, we match Eq.
(\ref{Rz}) with Eq. (\ref{limr}) at some intermediate point $
z=z_{m} $ which may be put as,
\begin{eqnarray}
J_{-}z_{m}&=&\frac{1}{3}R(1)+\frac{2}{3}R(1)z_{m}+\frac{BR(1)}{3
r_{+}^2}(1-z_{m}) \nonumber \\
&&+\frac{1}{2}\bigg[-\frac{4}{9}-\frac{2B}{9
r_{+}^2}+\frac{B^2}{18 r_{+}^4}-\frac{\phi'^2(1)}{18
r_{+}^2}\bigg]R(1)(1-z_{m})^2,
\end{eqnarray}
\begin{eqnarray}
J_{-}=\frac{2}{3}R(1)-\frac{BR(1)}{3 r_{+}^2}+
\bigg[\frac{4}{9}+\frac{2B}{9 r_{+}^2}-\frac{B^2}{18
r_{+}^4}+\frac{\phi'^2(1)}{18 r_{+}^2}\bigg]R(1)(1-z_{m}).
\end{eqnarray}
It is a matter of calculations to show that the above set of
equations, lead to the following quadratic equation for $ B $
\begin{eqnarray}\label{B^2}
B^2+4r_{+}^2\bigg(\frac{2+z_{m}^2}{1-z_{m}^2}\bigg)B+4r_{+}^4\bigg(\frac{1+2z_{m}^2}{1-z_{m}^2}\bigg)-\phi'^2(1)r_{+}^2=0.
\end{eqnarray}
Again, it is a matter of calculation to show that Eq. (\ref{B^2})
can by obtained by combining  Eq. (\ref{limr}) with condition $
J_+ =0$ at the asymptotic infinity $ z\rightarrow0$.
 Eq. (\ref{B^2}) has a solution of the form
\begin{eqnarray}\label{eq44}
B=\sqrt{\phi'^2(1)r_{+}^2+12r_{+}^4
\frac{(1+z_{m}^2+z_{m}^4)}{(1-z_{m}^2)^2}}-2
r_{+}^2\bigg(\frac{2+z_{m}^2}{1-z_{m}^2}\bigg).
\end{eqnarray}
We consider the case in which the value of the external magnetic
field ($B$) is very close to the upper critical value i.e, $ B\sim
B_{c} $. Thus, one can neglect all the quadratic terms in $ \psi
$. Under this condition, Eq. (\ref{phiz}) simplifies to
\begin{eqnarray}
\bigg(2+ b\frac{z^4}{r_{+}^2}\phi'^2(z)\bigg)\phi''(z)+\frac{4 b
z^3}{r_{+}^2} {\phi'^3(z)}=0.
\end{eqnarray}
Next, we integrate the above equation in the interval $ [1,z] $,
after using $ \phi'(1)<0 $ and the asymptotic boundary condition
for $ \phi $ given in Eq. (\ref{limphi}), we arrive at
\begin{eqnarray} \label{phip}
\phi'(z)=\frac{r_{+}(1-\sqrt{1+2b z^4 \lambda^2})}{bz^4 \lambda},
\end{eqnarray}
where $ \lambda={\rho}/{r_{+}^2} $. Expanding (\ref{phip}) for
small value of $b$ leads to the following expression at the
horizon ($z=1$),
\begin{eqnarray}\label{eq46}
\phi'(1)=-\frac{\rho}{r_{+}}.
\end{eqnarray}
From Eqs. (\ref{Tc}), (\ref{kappa}) and invoking the value of $
\tilde{\beta} $ from Eq. (\ref{eq27}), we have
\begin{eqnarray}\label{rho}
\rho=\frac{16 \pi^2 \tilde{\beta}}{9}T_{c}^2(0) \bigg[1+2b
\tilde{\beta}^2(1-z_{m})\bigg],
\end{eqnarray}
where $ T_{c}(0) $ is the critical temperature at zero magnetic
field. Combining Eqs. (\ref{eq44}), (\ref{eq46}) and (\ref{rho}),
we find the expression of the critical magnetic field as
\begin{eqnarray}
\frac{B_{c}}{T_{c}^2(0)}&=&\frac{16}{9}\pi^2
\Bigg{\{}\tilde{\beta}\sqrt{\bigg(1+2b
\tilde{\beta}^2(1-z_m)\bigg)^2+\frac{12(1+z_{m}^2+z_{m}^4)}{\tilde{\beta}^2(1-z_{m}^2)^2}\bigg(\frac{T}{T_{c}(0)}\bigg)^4}
-\frac{2(2+z_{m}^2)}{(1-z_{m}^2)}\bigg(\frac{T}{T_{c}(0)}\bigg)^2\Bigg{\}}
 \nonumber\\ &&+\mathcal{O}(b^2).
\end{eqnarray}
Considering the fact that the nonlinear parameter $b$ is small,
and following \cite{Lala,Roy}, we can write the above relation as
\begin{eqnarray}  \label{Bc}
\frac{B_{c}}{T_{c}^2(0)}&\simeq&\frac{16}{9}\pi^2\left(1+\frac{2b\tilde{\beta}^2(1-z_{m})}
{1+\frac{12(1+z_{m}^2+z_{m}^4)}{\tilde{\beta}^2(1-z_{m}^2)^2}\bigg(\frac{T}{T_{c}(0)}\bigg)^4}\right)\nonumber
\\ && \times \Bigg{\{}\tilde{\beta}\sqrt{1+\frac{12(1+z_{m}^2+z_{m}^4)}{\tilde{\beta}^2(1-z_{m}^2)^2}\bigg(\frac{T}{T_{c}(0)}\bigg)^4}
-\frac{2(2+z_{m}^2)}{(1-z_{m}^2)}\bigg(\frac{T}{T_{c}(0)}\bigg)^2\Bigg{\}}
  +\mathcal{O}(b^2). \nonumber\\
\end{eqnarray}
\begin{figure}
\centering\includegraphics[scale=.5]{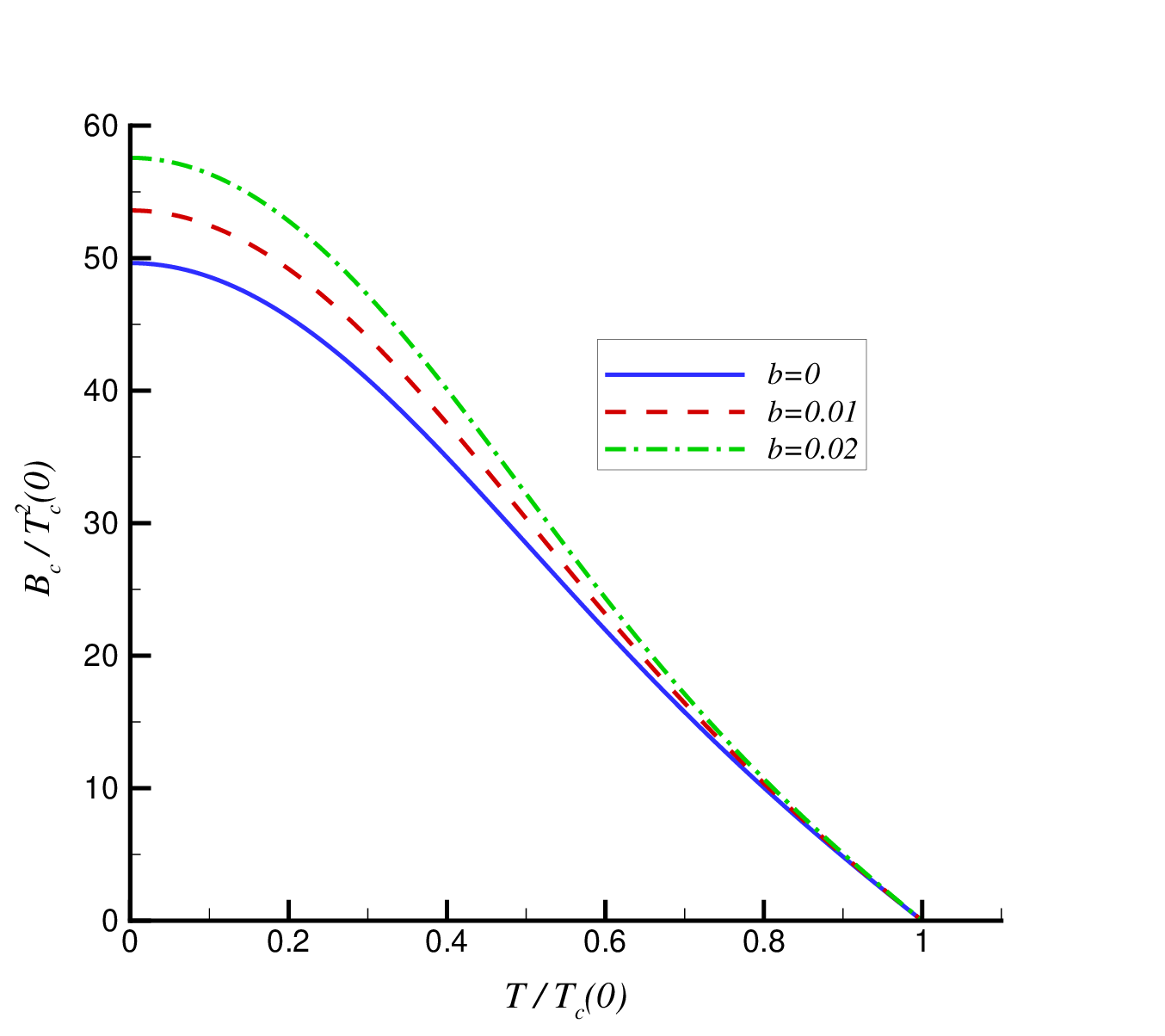}\caption{ The behaviour
of $ B_{c}/T_{c}^2(0)$ in terms of $T/T_{c}(0) $ at $ z_{m}=0.5 $
for different value of nonlinear parameters $b$.}\label{Fig1}
\end{figure}
\begin{figure}
\centering\includegraphics[scale=.5]{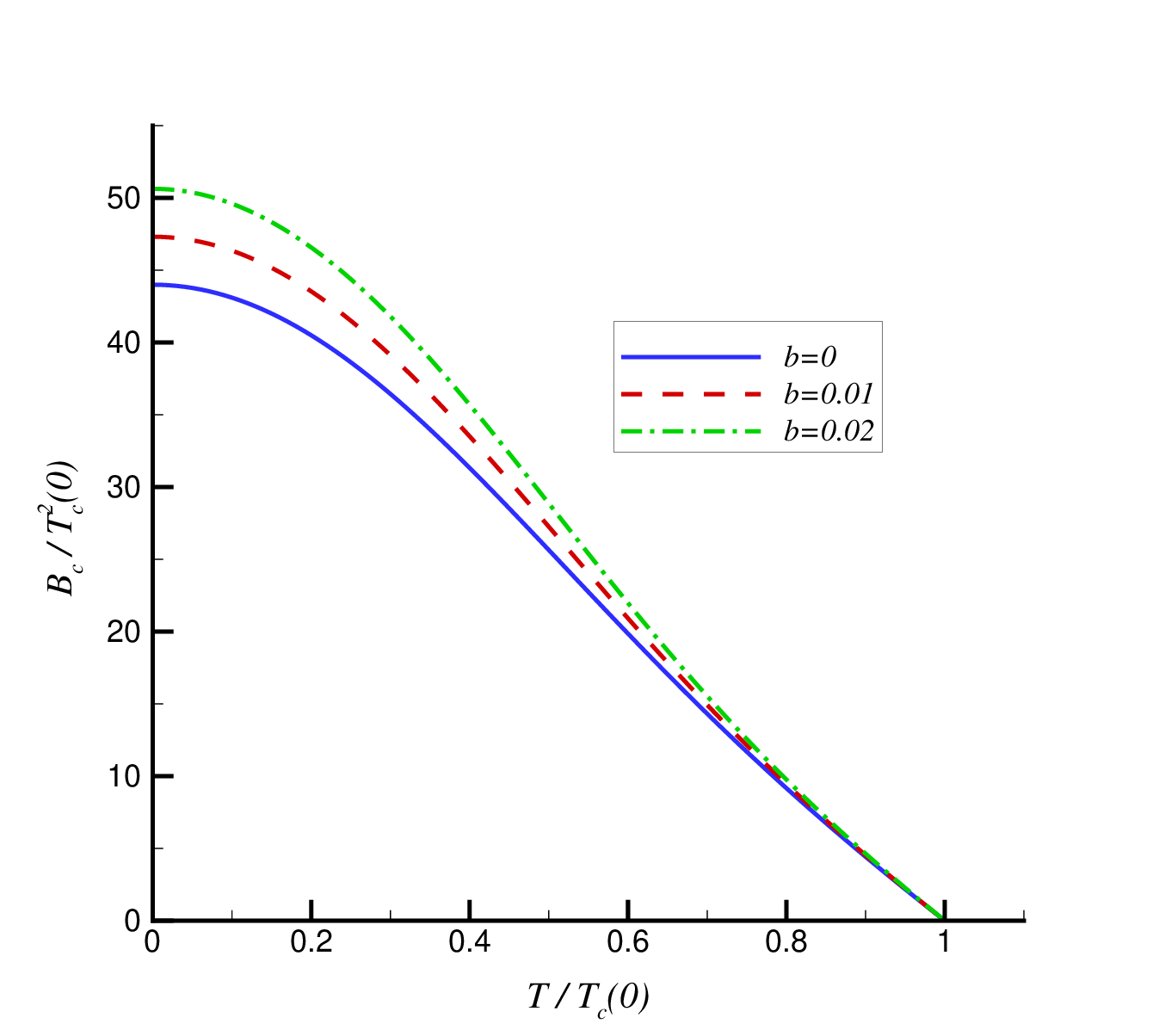}\caption{The behaviour
of $ B_{c}/T_{c}^2(0)$ in terms of $T/T_{c}(0) $ at $ z_{m}=0.4 $
for different value of nonlinear parameters $b$.}\label{Fig2}
\end{figure}
\begin{figure}
\centering\includegraphics[scale=.5]{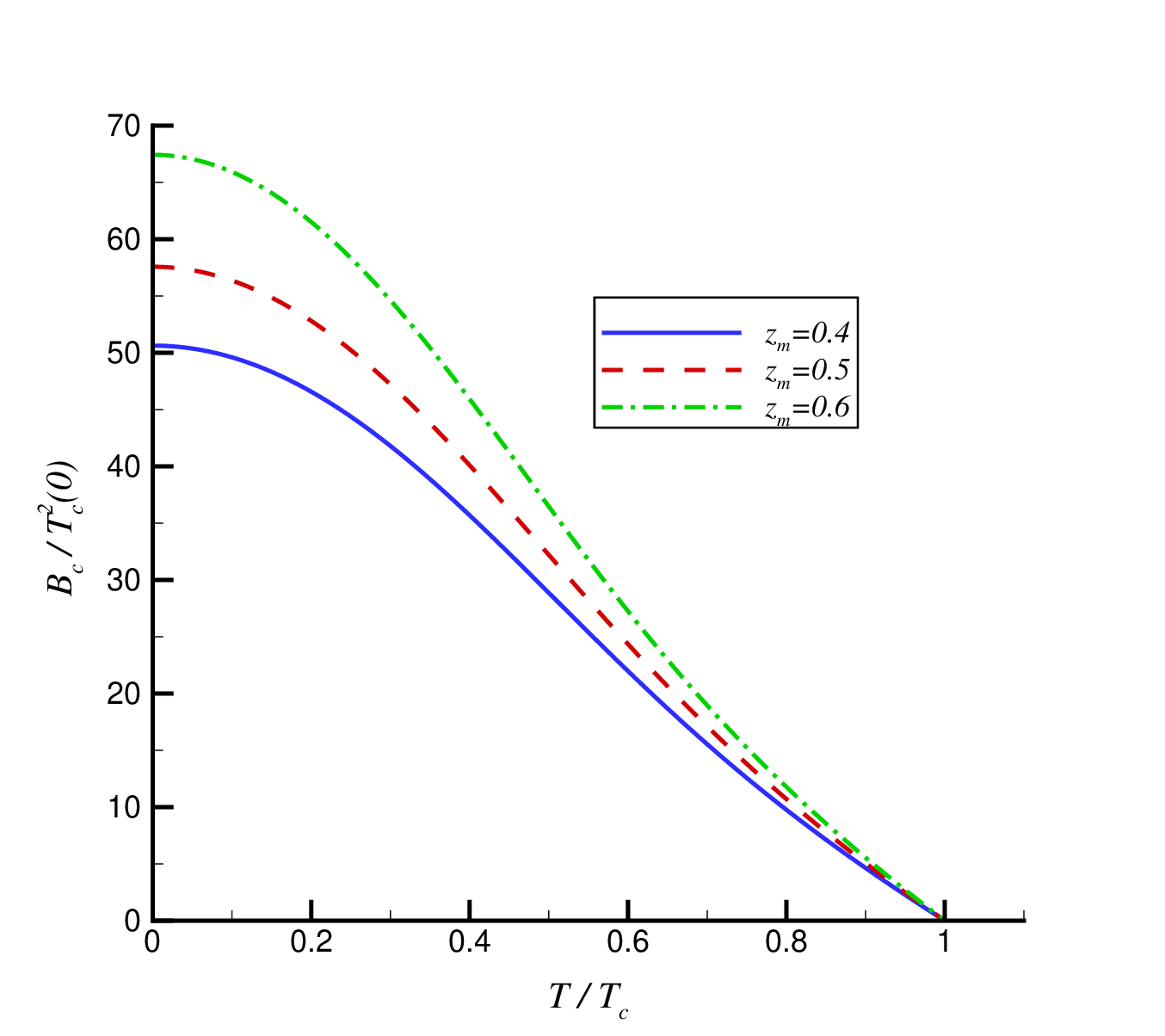}\caption{ The behaviour
of $ B_{c}/T_{c}^2(0)$ in terms of $T/T_{c}(0) $ for $ b=0.02 $
and different values of $z_m$.}\label{Fig3}
\end{figure}
The above result clearly reveals the dependence of the critical
magnetic field on the nonlinear parameter $b$. Note that this
result is valid for small values of the nonlinear parameter $b$.
In the limiting case where $ b\rightarrow0 $, the results of
holographic superconductor in Maxwell theory in the presence of an
external magnetic field are restored \cite{XH}. We have plotted
the behavior of the critical magnetic field in terms of
temperature in Figs. $1-3$. From Figs. $1$ and $2$ we  see that
$B_c$ is larger for LN holographic superconductor ($b\neq0$)
compared to the Maxwell case ($b=0$). Also, we find out that for
temperature smaller than the critical temperature, $T<T_c$, the
critical magnetic field increases with increasing $b$ and goes to
zero as $T\rightarrow T_c$, independent of the nonlinear parameter
$b$. Finally, from Fig. $3$, we observe that for a fixed value of
$b$, the critical temperature increases with increasing the
intermediate point $z_m$.
\section{Conclusions}
In this paper, based on the matching technique, we have
investigated various properties of the holographic $s$-wave
superconductors in an external magnetic field and in the presence
of logarithmic nonlinear electrodynamics (LNE). Considering the
probe limit in which the scalar and gauge field do not affect on
the background metric, we have presented a detailed analysis of
solving the coupled equations of motion for the scalar and the LNE
gauge field. We have obtained the relationship between the
critical temperature and the charge density. We kept the terms up
to the linear order in the nonlinear parameter $b$ and neglected
$\mathcal{O}(b^2)$ and higher order terms by assuming that $b$ is
small. We have summarized our result for the critical temperature
in table $1$. From this table, we understand that the critical
temperature for condensation, $T_{c}$, decreases with increasing
the values of the nonlinear parameter $b$. The variation of the
order operator with temperature, $<\mathcal{O}_{-}> \sim
\sqrt{1-\frac{T}{T_{c}}}$, exhibits a mean field behavior with
critical exponent ${1}/{2}$. It is worth mentioning that the
expressions for the critical temperature and the condensation
values were obtained up to the linear order in $b$.

We have also disclosed the effects of an external magnetic field
on the holographic superconductor in the presence of LNE. We have
found an expression for the critical magnetic field $B_c$ in terms
of the temperature. In the limiting case where $ b\rightarrow0 $,
all results reduce to those of the holographic superconductor with
magnetic field in Maxwell theory \cite{XH}. We have investigated
the behavior of the critical magnetic field in terms of the
temperature by plotting $B_{c}/T_{c}^2(0)$ versus $T/T_{c}(0)$. We
observed that for $T<T_c$, the critical magnetic field increases
with increasing the nonlinear parameter $b$ and goes to zero at
$T=T_c$ independent of the value of $b$ (see Figs. $1$ and $2$).
This implies that the nonlinear electrodynamics makes condensation
harder. Finally, we found out that for a fixed value of $b$, the
critical magnetic field increases as the intermediate point $z_m$
approaches to the boundary ($z=1$) (Fig $3$).

Note that in this paper, we have limited our study to the case
where the scalar and gauge fields do not back react on the metric
background. It is interesting to generalize this study by
exploring the effects of external magnetic field as well as LNE on
the holographic superconductor away from the probe limit. One may
also consider the case with exponential nonlinear electrodynamics.
These issues are currently under investigation and the results
will be reported subsequently.

\acknowledgments{We thank the Research Council of Shiraz
University. This work has been supported financially by Research
Institute for Astronomy and Astrophysics of Maragha (RIAAM),
Iran.}

\end{document}